# Results of the 2015 Workshop on Asteroid Simulants

Philip T. Metzger[1], Daniel T. Britt[2], Stephen D. Covey[3], and John S. Lewis[4]

[1]University of Central Florida, Florida Space Institute, 12354 Research Parkway, Suite 214, Orlando, FL 32826-0650; PH 407-823-5540; email: Philip.Metzger@ucf.edu
[2]University of Central Florida, Department of Physics, 4111 Libra Drive, Physical Sciences Bldg. 430, Orlando, FL 32816-2385; PH 407-823-2600; email: dbritt@ucf.edu
[3]Deep Space Industries, Inc., 13300 Tanja King Blvd, #408, Orlando, FL 32828; PH 904-662-0550; email: Stephen.Covey@DeepSpaceIndustries.com
[4]Deep Space Industries, Inc., P.O. Box 67, Moffett Field, CA 94035; PH 855-855-7755; email: John.Lewis@DeepSpaceIndustries.com

## ABSTRACT

The first asteroid simulants workshop was held in late 2015. These materials are needed for tests of technologies and mission operational concepts, for training astronauts, for medical studies, and a variety of other purposes. The new program is based on lessons learned from the earlier lunar simulants program. It aims to deliver families of simulants for major spectral classes of asteroids both in cobble and regolith form, beginning with one type of carbonaceous chondrite and rapidly expanding to provide four to six more asteroid classes. These simulants will replicate a selected list of asteroid properties, but not all known properties, in order to serve the greatest number of users at an affordable price. They will be benchmarked by a variety of data sets including laboratory analysis of meteorites, observation of bolides, remote sensing of asteroids, data from asteroid missions, and scientific modeling. A variety of laboratory tests will verify the as-manufactured simulants are accurately and repeatedly providing the specified characteristics.

## INTRODUCTION

The first asteroid simulants workshop was held October 6-7, 2015 at the offices of the Florida Space Institute (FSI), part of the University of Central Florida. It was co-sponsored by Deep Space Industries, FSI, and the Center for Lunar and Asteroid Surface Science (CLASS), a node of NASA's Solar System Exploration Research Virtual Institute (SSERVI). The attendees reviewed the history of lunar soil simulants in order to avoid the problems that were encountered in the lunar program and to adopt its best practices. Then they identified needs for asteroid simulants and determined their requirements. Finally, they developed a strategy including which types of simulants to make first, how to validate them, and how they should be stored and distributed. This paper reports on the proceedings.

## BACKGROUND: LUNAR SIMULANTS

The Apollo program demonstrated the challenges in designing hardware to work with extraterrestrial regolith, and the lessons are equally important for asteroid simulants. For example, the drive tubes to obtain core samples of lunar regolith needed to be redesigned after Apollo 11 and again after Apollo 14 because of the difficulty driving the tube into highly frictional and compacted lunar regolith, and because the sample that was driven into the tube became unacceptably disturbed by the geometry of the tube. The development cycle for space technology will become more successful and less expensive if high fidelity simulants are available earlier.

The low fidelity simulant used for drive tube testing was a mixture of kaolinite clay and League City (Texas) sand. After the Apollo program, several improved lunar soil simulants were developed, including Minnesota Lunar Simulant-1 (MLS-1) [Weiblen and Gordon, 1988; Weiblen et al, 1990] and Johnson Space Center-1 (JSC-1) [Willman et al, 1995]. JSC-1 was designed for geotechnical purposes and to a lesser extent chemistry of lunar mare soil, although its chemistry was really not typical of mare soil [Taylor and Liu, 2010]. The simulant was often used, perhaps inappropriately, in tests that needed a better chemical simulant. Over time additional simulants were developed by various users, everyone according to his or her unadvised opinion. This is because the vast majority of users have neither the time nor background to develop regolith consistent with the details of lunar geology. Many tests were performed with incorrect simulants resulting in wasted time, or worse, deceptive results that in the extreme of a spaceflight program could have tragic consequences [LEAG, 2010].

To rectify this, the NASA Marshall Spaceflight Center (MSFC) was designated to manage lunar simulants and they developed the 5-step approach given in Table 1. (This is the process we will follow in this asteroid simulants project in close collaboration with the NASA team.) Having MSFC act as the central clearinghouse following this rigorous process enabled NASA to “obtain better simulants, with rigorous specifications and performance, and at lower expense” [McLemore, 2014]. Unfortunately, many researchers still set about on their own, disregarding the structure NASA had set in place, and so many lunar simulants were developed without working with the NASA MSFC team. As a result, by 2010 more than 30 simulants had been developed by various groups in the U.S. and overseas [LEAG, 2010]. The result was (1) many poorly designed simulants that had incorrect properties and (2) the rampant misuse of well-designed simulants by expecting them to have properties that they weren’t designed for [Taylor and Liu, 2010]. However, in the midst of that chaos the NASA team with its contractors successfully re-created and characterized the JSC-1 simulant, now as JSC-1A, and created the high-fidelity chemical/mechanical lunar highlands simulant series NU-LHT. These pedigreed simulants along with the careful means to measure them gave NASA the tools it needed for writing contracts to develop hardware such that it could work with lunar soil. It also gave the more careful technologists who did not misuse the simulants the ability to compare results with one another and with lunar soil. With more progress developing pedigreed simulants and demonstrating the benefits of their use in publications, the tide of poor lunar simulant engineering outside of the NASA-led team may eventually be curbed.

**Table 1. Five Step Approach to Developing Simulants** [McLemore, 2014]

| |
|---|
| 1. Development of the necessary simulant requirements based on the appropriate standards |
| 2. Development of a method by which to compare simulants to a reference |
| 3. Selection and measurement of the appropriate reference materials. |
| 4. Development and demonstration of simulant development and process control techniques |
| 5. Selection of suitable simulant feed stocks |

If the lunar community could have the ideal soil simulant, it would many properties in common with actual lunar soil. The lunar community created a list of 32 desired properties and organized them into ten categories [McKay and Blacic, 2002]. Some of these such as agglutinates are not relevant or significant in the asteroid case, while additional properties will need to be added. For example, with asteroids we may desire to replicate the volatile liberation temperature for water and organic matter as well as the chemistry of the organic matter.

It turns out that it is far too difficult – perhaps impossible – and far too expensive to create a simulant with all of the desired properties. Extraterrestrial regolith is too exotic compared to terrestrial materials and too complex to engineer with perfect fidelity. Thus, choices had to be made about which properties would be simulated, and the users did not all have the same requirements. Some wanted simulants for different parts of the Moon (just as it will be with asteroid simulants and different types of asteroids); some wanted broad-use simulants whereas others had very specific needs; some needed higher fidelity than others; some needed fine dust while others needed the coarser fraction or even cobbles; and they needed simulants that had undergone different degrees of weathering on the lunar surface [Stoeser 2009]. Thus, it became clear that one simulant would not work for all needs, and this is one of the main factors leading to the misunderstanding and misuse of lunar simulants. Many users thought that a simulant was a simulant was a simulant. It turns out that with the simulants developed so-far, the community focused upon just three properties or groups of properties: the particle size distribution, the mechanical properties, and the chemical properties [McKay, 2009]. Some users amended simulants in various ways or created ad hoc simulants for specialized tests.

A rigorous and cost-effective way to deal with this situation is to develop *families* of simulants with roots and branches, representing the basic simulant and its variants for specific needs. The NASA simulants program developed generally two such families, the JSC and NU-LHT series. The JSC-1 simulant was re-constituted as JSC-1A [Gustafson, 2009] and a series of its adaptations were developed, including JSC-1Af for lunar dust, JSC-1Ac for coarse particles, and a version of JSC-1A that included agglutinated particles made via plasma torch processing [Gustafson et al, 2008].

These simulants somewhat represent the chemistry of lunar maria-type regolith. For the lunar highlands, the very high fidelity NU-LHT series of simulants was developed to meet most chemical and geotechnical needs. They have a mixture of minerals to represent the lunar highlands, plus simulated breccias as well as glass including pseudo-agglutinates that give it high-fidelity mechanical behavior [Stoeser et al, 2008; Stoeser et al, 2010]. In addition to the US-developed simulants, a number of international simulants were assessed by the NASA simulants team. A Canadian team developed the simulant OB-1 for lunar drilling tests [Battler et al, 2006; Richard et al, 2007; Battler and Spray, 2009] and a follow-on version, Chenobi [Electric Vehicle Controllers, 2009]. FJS-1 and MKS-1 were developed by the Japanese space agency [Jiang et al, 2011] and NAO-1, CAS-1, and TJ-1, were developed by different teams in China [Li et al, 2009; Zheng et al, 2009; Jiang et al, 2011].

The NASA MSFC team developed a Figure of Merit (FoM) to compare how well the various lunar simulants match a particular lunar soil [Rickman et al, 2010]. Based on their analysis of simulants available at that time, the Lunar Regolith Simulant User's Guide including a "Fit-to-Use" table was published to help technologists and scientists select which simulant to use for particular activities [Rickman et al, 2010]. This guide comprised analyses of MLS, the JSC-1 series, the NU-LHT series, OB-1, Chenobi, and FJS.

## PRIOR USE OF ASTEROID SIMULANTS

A literature search (as exhaustive as possible) found that, to-date, various ad hoc simulants are being used for a wide variety of asteroid studies for technology and for pure science. Housen [1992] used a mixture of 50% basalt fragments, 24% fly ash, 20% iron grit, and 6% water to simulate asteroid regolith for cratering ejecta experiments. Fujiwara et al [2000] used "various kind of rocks, sand, and artificial materials like bricks" while Yano et al [2002] used glass beads and lunar soil simulant to study asteroid sampling via projectile impact. Sears et al [2002] studied the formation of smooth regolith ponds on asteroids by using Martian regolith simulant JSC-Mars-1 and with mixtures of sand plus iron grains. Sandel et al [2006] used "meteorite simulants" for impact experiments to study collisional disruption and resulting fragment distribution from asteroids. Izenberg, and Barnouin-Jha [2006] used playground sand with embedded cobbles to simulate asteroid regolith to study how impacts affect the morphology and vertical layering of asteroids. Makabe [2008] used a simulant of a C-type asteroid to study projectile impact in the Hayabusa-2 mission for capturing samples. Güttler et al [2012] studied crater formation on asteroid surfaces using spherical glass beads. Barucci et al [2012] used a lunar regolith simulant and "many simulants" to test an asteroid sampling mechanism. Durda et al [2012, 2013, 2014] studied the morphology of asteroid surfaces using lunar soil simulant JSC-1A, glass microspheres, and bread flour. Bernold [2013] performed asteroid mining and conveying experiments using lunar regolith simulants. Crane et al [2013] used shaving from a steel bar to simulate a Tholen Type M asteroid regolith for thermal inertia tests. Murdoch et al [2013] studied the strength properties of asteroid regolith in microgravity using spherical soda-lime glass beads. Backes et al [2014] used floral foam and "a variety of simulants" both hard and soft to represent the surface of a comet. This survey shows that, at the present, the materials chosen for asteroid simulants are generally low-fidelity or inappropriate and lack the commonality and standards that would promote comparing results. (Not included in this list are the many spectroscopic studies of terrestrial analog materials that inform remote observation and calibrate spacecraft instruments, nor space weathering experiments or the like.) This review also demonstrates the wide range of uses for simulants, and it indicates the need for leadership in producing easily accessible, pedigreed simulants (along with sample preparation procedures) in appropriate levels of fidelity and cost.

## NEEDS FOR ASTEROID SIMULANT

Simulants will be needed to support the development of space missions. These include:

- Japan's Hayabusa-2 mission, which has NASA-funded collaborators and which will perform explosive cratering and sample return of the asteroid's surface
- NASA's Origins-Spectral Interpretation-Resource Identification-Security-Regolith Explorer (OSIRIS-REx) mission, which will use a burst of gas to collect asteroid regolith for sample return
- NASA's Asteroid Redirect Mission (ARM), both the robotic portion and the crew exploration portion
- Space mining by commercial companies
- Future missions for planetary defense

Science missions need asteroid simulants to develop sampling devices, to test digging methods, to test sensors and dust mitigation, etc. Simulants are used during the mission to compare to results that are observed at the destination, to interpret the results and decide upon a course of action. For example, when a Mars Exploration Rover had trouble driving, simulants were used to determine the best wheel motions to get un-stuck or to avoid getting stuck. The robotic portion of the ARM will require simulants to test asteroid grappling devices and/or boulder extraction from an asteroid. The human mission to visit the returned asteroid will need simulants to develop crew tools for studying and sampling the asteroid.

NASA's technology development program will require asteroid simulants. These are predominantly in the following Technology Areas: TA04 Robotics, Tele-Robotics and Autonomous Systems; TA06 Human Health, Life Support, and Habitation Systems; TA07 Human Exploration and Development of Space; TA08 Science Instruments, Observations and Sensor Systems; and TA09 Entry, Descent and Landing; although some applications exist in the other Technology Areas, as well. Some examples include: mobility testing of robotics on asteroid regolith; human health when exposed to asteroid dust and organic molecules; mining and processing of resources from asteroids; instruments to study asteroids; and propulsion systems that will not overly disturb asteroid regolith during proximity operations.

In addition to the NASA need for asteroid simulants, there is a newfound commercial need for them. Multiple companies, Deep Space Industries included, have announced plans to mine asteroids for volatiles and other components. In the early stages of asteroid mining, the low-hanging fruit is water for life support and propulsion. Extracting $CO_2$ in addition to $H_2O$ enables the production of soft cryogens including methane and liquid oxygen, and even the production of asteroid-derived storable propellants such as hydrogen peroxide and dimethyl ether. The development of these commercial applications requires simulants for anchoring, excavation, and volatiles extraction. Future commercial applications include similar developments leading toward the extraction of structural metals and silicon for solar cells.

Planetary defense is another potential user of asteroid simulants. Techniques must be developed and tested that enable deflection of asteroids with impending impacts on the Earth. In some of these scenarios, asteroid simulants will be vital in developing the kinetic impact vehicles, landers and thrusters to redirect asteroids.

Lunar simulants have been used in studying the health effects of lunar dust. Similar research must be performed for asteroids. The concerns include inhalation, dermal toxicity, ocular toxicity, and dissolution into the body. In the lungs, dust can cause edema, fibrosis, inflammation, and possibly cancer [Khan-Mayberry, 2009]. The particle sizes less than 10 microns are considered respirable, so simulants comprising that size range are necessary. The Lunar Airborne Dust Toxicity Assessment Group decided that it was necessary to test a variety of lunar dusts. [Khan-Mayberry, 2009]. Varieties include mature vs. immature and highlands vs. mare. The need for variety is more acute with the greater variety of asteroids.

## FAMILIES OF ASTEROID SIMULANT

Because there are many spectral classes of asteroids (and many types of meteorites that are samples of those asteroids), there need to be many types of simulant. The workshop decided it is best to develop one type of simulant first and after validating it to develop four to six more. The workshop decided that perhaps a Carbonaceous CI simulant would be a good choice for the first, followed by CM, C2, CV, L Ordinary, LL Ordinary, H Chondrite, Iron, Enstatite Chondrite, and Basaltic Chondrite types (not listed in order of preference, which is yet to be determined).

Each of these simulants will be a “root” simulant. More specialized versions, or “branches”, can be developed to meet specific user needs. For example, a version with higher fidelity organic content may be needed for medical tests whereas most users do not want that because it is carcinogenic.

Simulants can be provided in two basic forms. The powderized mineral constituents can be bonded together to form competent cobbles or even bounders, or it can be provided as loose regolith. For the regolith form, there may be different versions based on the particle size distribution, including standard, fine, and coarse variations.

The cobble form is manufactured by grinding the mineral constituents to the desired textures, mixing them, wetting, and drying. Preliminary work shows the clay component binds these powders remarkably well. The details of wetting and drying determine the mechanical strength of the resulting cobbles. For regolith simulant of the highest fidelity, it would be necessary to form cobbles in this manner and then re-grind them so the individual regolith grains are themselves lithic fragments of multiple minerals. However, it is not envisioned that we shall provide such high fidelity regolith at this time. Instead, it will be provided as mixtures of monomineralic grains (the mixed powders prior to wetting and drying). Users may procure and crush the cobbles if they need higher fidelity.

## PROPERTIES OF ASTEROID SIMULANT

The workshop compiled an extensive list of asteroid properties that may (or may not) be desired for asteroid simulants. These are listed below. The attendees decided which parameters were so important that the simulant design must be controlled by them. These are listed in bold with an asterisk. The attendees also decided which parameters are important to measure and report to the user community although they are not control parameters. These are listed in italic with an asterisk.

**Table 2. Properties of Asteroid Simulants**

| |
|---|
| 1. Grain properties<br>a. **Size distribution*** (data from power law observations)<br>i. **Mean particle size***<br>ii. Broadness of size distribution<br>iii. Coefficient of curvature<br>iv. Coefficient of uniformity<br>v. Internal erodibility<br>b. *Particle Shapes distribution**<br>c. Morphology<br>d. Specific surface area<br>e. Intra-grain porosity |
| 2. Electrostatic properties (depends critically on the environment and is hard to replicate in a laboratory) |
| 3. **Magnetic Properties*** |
| 4. Geomechanical Properties<br>a. Fatigue<br>b. **Tensile Strength***<br>c. **Compressive Strength***<br>d. **Shear Strength***<br>e. Grain Hardness (hardness indexes)<br>f. Surface friction<br>g. Abrasivity (for tool development)<br>h. Flexural Strength-bending resistance<br>i. Fracture properties, friability<br>j. Impact resistance<br>k. Rheology<br>l. *Angle of Repose**<br>m. *Regolith Internal Friction**<br>n. *Regolith Cohesion**<br>o. Adhesion (depends on tool material, too)<br>p. Compressibility of regolith<br>q. Compactibility of regolith (index test, like Proctor Compaction) |
| 5. Optical properties<br>a. *Albedo**<br>b. *Reflectance spectrum**<br>c. Absorption<br>d. Thermal emissivity |
| 6. Aerodynamic properties<br>a. Gas erodibility (rocket exhaust)<br>b. Particles' coefficient of drag |
| 7. Physical Properties<br>a. Thermal properties (derived properties from mineralogy, texture, and volatile content)<br>i. heat capacity<br>ii. Conductance<br>iii. thermal cracking behavior<br>iv. Emissivity<br>b. **Bulk density of rocks***<br>c. Particle density<br>d. **Porosity of rocks***<br>e. Surface area<br>f. Permeability of rocks<br>g. Permeability of regolith as a function of porosity/compaction<br>h. Bulk density of regolith as a function of porosity/compaction |
| 8. Geochemical properties<br>a. Bulk chemistry (derived property of the composition)<br>b. **Mineralogy***<br>c. Siderophile elements in Iron simulants<br>d. Modal Composition<br>e. Isotopic ratios<br>f. Organic content<br>i. C-to-H ratio (aliphatic vs aromatic)<br>ii. Toxicity<br>iii. Sulphur and Nitrogen content of the organic matter |
| 9. Chemical reactivity<br>a. From surface damage<br>b. As volatile /soluble minerals<br>c. **Absorptive capacity for volatiles*** |
| 10. Texture<br>a. Homogeneity and isotropy of texture<br>b. Chondrules |
| 11. Volatiles<br>a. Volatiles content<br>b. **Water***<br>c. **Organics***<br>d. **Sulphur compounds***<br>e. **Release pattern***<br>i. **thermal and/or vacuum release***<br>ii. **chemisorbed, physisorbed patterns***<br>f. Implanted solar wind particles |

## VALIDATION OF ASTEROID SIMULANTS

Next, the attendees decided how asteroid simulant should be tested. There are two purposes in this. First, it benchmarks the simulant's properties as-designed to real asteroids so that users will understand both the value and limitations of simulant in lieu of real space materials. Users can then design meaningful applications and interpret the results in terms of the space materials. Second, it validates the accuracy and repeatability of the simulant's properties as-manufactured so the user can be sure the simulant is the same as when the original batch was benchmarked. Benchmarking will make use of the following data sources:

- Meteorites
  - Laboratory measurements
  - Bolide observations to determine compressive strengths
- Asteroids
  - Ground-based observations by radar and thermal infrared
  - Spacecraft imagery and sample return

- Modeling
  - Interparticle cohesion
  - Depletion of fine particles and buildup of surface lags
  - Particle size distribution to match remote sensing
  - Theories of formation

The data from these sources will be integrated to develop simulant testing requirements such as particle size distribution and compressive strength. Several types of meteorites will be needed for laboratory measurements because several types of asteroid simulants will be developed representing the different classes of asteroids (CI, CM, H Ordinary, etc). For documented specificity, the laboratory measurements will be performed on selected meteorites in each class. While the list is expected to evolve, the workshop decided that appropriate reference meteorites could include the following: CI – Orgueil; CM – Murchison; C2 – Tagish Lake; CV – Allende; Iron – Gibeon. The attendees decided that reference meteorites should also be specified for the following types: L Ordinary Chondrite; LL Ordinary Chondrite; H Ordinary Chondrite; Enstatite Chondrite; and Basaltic Achondrite. After benchmarking the prototype of each simulant class with these data sources, batches of manufactured simulant will be tested using statistically relevant samples to verify the manufacturing processes are adequately controlled. The tests will check each of the control parameters, including bulk density measurements via immersion in a wetting fluid, particle size distribution via sieving, mechanical hardness using standard engineering test equipment, thermogravimetric tests of water release as a function of temperature, mass spectrometry of the released volatiles as a function of temperature, and weighing on a Faraday balance for magnetic susceptibility.

## CONCLUSION

The first asteroid simulant workshop made important progress in defining a proactive, well-documented program. By basing this program on methods that NASA developed for lunar simulants, the asteroid simulants program will attempt from the beginning to avoid misunderstandings and misuse of simulants that occurred in the lunar community. This program will provide the asteroid researchers and technologists with several families of consistent, well-documented asteroid simulants that are pedigreed through benchmarking against space materials. This program will help ensure comparability of tests between the users, higher quality tests, and cost savings since every project will not need to develop simulants on its own.

## ACKNOWLEDGEMENT

The authors gratefully acknowledge the assistance of Dr. Jim Mantovani of NASA's Kennedy Space Center and support for this work from NASA's Small Business Innovative Research (SBIR) 2015 program, Phase 1, subtopic H1.01, "Regolith ISRU for Mission Consumable Production," contract NNX15CK10P.